\documentclass[useAMS,usenatbib]{mn2e}

\usepackage[activeacute,swedish,english]{babel}
\usepackage{graphicx}
\usepackage[T1]{fontenc} 
\usepackage{latexsym}
\usepackage{verbatim}
\usepackage{amsmath}
\usepackage{subfigure}

\def\gtappeq{\mathrel{ \rlap{\raise.5ex\hbox{$>$}}
                      {\lower.5ex\hbox{$\sim$}}  } }

\def\leappeq{\mathrel{ \rlap{\raise.5ex\hbox{$<$}}
                      {\lower.5ex\hbox{$\sim$}}  } }

\title[Two new accreting, pulsating WDs: J1457 and BW Scl]{Two new accreting, pulsating white dwarfs: \newline SDSS J1457+51 and BW Sculptoris} 
\author[Uthas et al.]{Helena Uthas$^{1}$\thanks{E-mail:
helena@astro.columbia.edu}, Joseph Patterson$^{1}$, Jonathan Kemp$^{2}$, Christian Knigge$^{3}$, \newauthor  Berto Monard$^{4}$, Robert Rea$^{5}$, Greg Bolt$^{6}$, Jennie McCormick$^{7}$, \newauthor Grant Christie$^{8}$, Alon Retter$^{9}$ and Alex Liu$^{10}$ \\
$^{1}$ Columbia University, Department of Astronomy, New York, NY, USA\\ 
$^{2}$ Gemini Observatory, Northern Operations Center, 670 North A$`$ahoku Place, Hilo, HI, USA\\ 
$^{3}$University of Southampton, Department of Physics and Astronomy,
  Highfield, Southampton SO17 1BJ, UK\\ 
$^{4}$ CBA (Pretoria), Post Office Box 11426, Tiegerpoort 0056, South Africa\\ 
$^{5}$ CBA (Nelson), 8 Regent Lane, Richmond, Nelson, New Zealand\\ 
$^{6}$ CBA (Perth), 295 Camberwarra Drive, Craigie, Western Australia 6025, Australia\\ 
$^{7}$ CBA (Pakuranga), Farm Cove Observatory, Pakuranga, New Zealand\\ 
$^{8}$ CBA (Auckland), New Zealand \\ 
$^{9}$ CBA (Shoham), 86a/6 Hamaccabim St., PO Box 4264, Shoham, 60850,
Israel \\ 
$^{10}$ CBA (Exmouth) Norcape Observatory, P.O. Box 300, Exmouth, 6707, Australia}

\begin{document}

\date{Accepted 2011 October 20. Received 2011 October 19; in original form 2011 March 7} 

\pagerange{\pageref{firstpage}--\pageref{lastpage}} \pubyear{year}

\maketitle

\label{firstpage}

\begin{abstract}
We report the discovery of rapid periodic signals in the light curves of two cataclysmic variables with prominent white-dwarf components in their spectra, SDSS J1457+51 and BW Sculptoris. These stars  therefore appear to be new members of the GW Lib class of variable star, in which the fast periodic (and non-commensurate with the orbital period) signals are believed to arise from non-radial pulsations in the underlying white dwarf. The power spectra of both stars show complex signals with primary periods near 10 and 20 minutes. These signals change in frequency by a few percent on a timescale of weeks or less, and probably contain an internal fine structure unresolved by our observations. We also detect double-humped waves signifying the underlying orbital periods, near 78 minutes for both stars.

In addition, BW Scl shows a transient but powerful signal with a period near 87 minutes, a quiescent superhump. The $\sim$ 11$\%$ excess over the orbital period is difficult to understand, and may arise from an eccentric instability near the 2:1 resonance in the accretion disc.

\end{abstract}


\section{Introduction}

Non-radial pulsations (NRPs) are commonly found in isolated white dwarfs (WDs) of DA type, so called ZZ Ceti stars. These stars have hydrogen-rich atmospheres, and pulsations occur as the WD cools and passed through a phase of pulsational instability, detected mainly as g-modes (\citealt{2006AJ....132..831G}).         

During the last decennium, similar signals, generally interpreted as non-radial WD pulsations, have also been detected in faint cataclysmic variables (CVs). A CV is a close binary system where a late-type main-sequence star looses mass to a primary white dwarf. The first CV proposed to harbour a pulsating white dwarf was GW Librae.~\cite{1998IAUS..185..321W} found rapid, periodic, and non-commensurate signals in its light curve, suggesting non-radial pulsations of the underlying white dwarf. In most CVs, the accretion energy tends to dominate the luminosity, and the white dwarf itself, shining with $M_{V} \sim$ 10 -- 13, is seldom seen. However, for some of the most intrinsically faint CVs, spectroscopy and time-series photometry can reveal signatures of the underlying white dwarf, such as broad absorption features in the spectrum, sharp eclipses, and sometimes non-radial pulsations in the light curve. These signals have now been detected in about a dozen CVs, all quiescent systems of low luminosity. Here, we call these systems \emph{GW Lib stars}, after the first discovery.~\cite{2010ApJ...710...64S} and~\cite{2009JPhCS.172a2069M} present recent reviews of this group of stars. 

Accreting WDs are different from isolated ones since they are being exposed to mass transfer, giving them atmospheres of solar-composition. The white dwarfs in CVs are therefore hotter and are also found to be spinning faster compared to isolated ones (\citealt{2009ASPC..404..229S}). Studying these systems will provide important information of how the process of accretion is affecting the evolution of the white dwarf. In isolated WDs, pulsations are only observed in stars with temperatures located within a so-called \emph{instability strip} in the $\log g$ -- $T_{\text{eff}}$ plane, spanning the temperature range $T_{\text{eff}}$ = 10900 K -- 12200 K (see Figure 3 of~\citealt{2006AJ....132..831G}). However, there is no clear instability strip for the GW Lib stars (see Figure 13 of~\citealt{2010ApJ...710...64S}), and pulsations are found in systems with WD effective temperatures up to at least 15000 K. 

Our theoretical understanding of the mechanism for exciting non-radial pulsations in CVs is still limited. The observed pulse amplitudes are quite variable, and in some ZZ Ceti stars this is known to be the result of the beating of two signals closely spaced in frequency, a classic and highly informative signature of non-radial pulsations. But the GW Lib stars have not yet clearly revealed this kind of behaviour (although a hint of it emerged in the V386 Ser campaign reported by~\citealt{2010ApJ...714.1702M}).  

We here present time-series photometry of two more CVs which are probably members of the GW Lib class. Both systems have very low accretion luminosity, and show signatures from the white dwarf in their optical spectra. Also, both systems show double-humped orbital signals, and non-commensurate periodic signals, suggesting non-radial pulsations. One is SDSS J1457+51 which has a photometric wave suggesting an orbital period of $77.885 \pm 0.007$ minutes. The other is BW Sculptoris, with $P_{\text{orb}} = 78.22639 \pm 0.00003$ minutes. Both stars show main pulsations near 10 and 20 minutes. These rapid signals drift slightly in frequency, and may consist of several, finely spaced components. BW Sculptoris also shows a remarkable photometric variation at 87 minutes, which could be explained as a \emph{quiescent superhump}, possibly arising from a 2:1 orbital resonance in the accretion disc.

\section{SDSS J1457+51}

SDSS J1457+51 (hereafter J1457) was first identified in the Sloan Digital Sky Survey by~\cite{2005AJ....129.2386S}. They obtained spectroscopy that showed the broad absorption characteristics of a white dwarf, indicating a system of low accretion rate. The source was found to be faint ($g \approx 19.5$). Due to the double-peaked nature of the emission lines, they suggested the system to be of high inclination.


\subsection{Observations, Data Reduction and Analysis} \label{obs}

Time-resolved photometry of J1457 was obtained with the 1.3 m and 2.4 m MDM telescopes at the Kitt Peak observatory, Arizona, during April and May 2010. The star was observed during 14 nights in total, spread over 47 days. With all data coming from the same terrestrial longitude, we were not immune from aliasing problems and therefore strove to obtain the longest possible nightly time series (generally $\sim$ 4 - 7 hours). Weather conditions such as clouds, snow and full moon prevented us from obtaining times series over more than 4 consecutive nights at a time. The time resolution was in the range of 20 s -- 30 s. Table~\ref{tab:obs1} presents a log of the observations. A clear filter with a blue cutoff was used to minimise differential-extinction effects and allow for a good throughput.

The data reduction was done in real time during the observations, using standard \textsc{Iraf} routines. The data consisted of differential photometry with respect to the field star USNO A2.0:1350-08528847. The search for periodic signals was initially done for single nights separately, and Lomb-Scargle periodograms (\citealt{1982ApJ...263..835S}) were constructed. Formal flux errors were rescaled so that the $\chi^{2}_{\nu} \approx$ 1. This was done by fitting a fake light curve to the original data, composed of multiple sine waves with periods corresponding to the strongest signals found in the single-night power spectrum. Monte-Carlo simulations were performed on every peak of interest in the power spectrum to find the period and its error. In this method, the peak errors are found by randomly re-distributing the points in the light curve within their errors, a repeated number of times, and constructing a Lomb-Scargle periodogram each time. The 1-$\sigma$ error is then found by fitting a Gaussian to the output distribution of the peaks found in the periodograms. Data from several nights were then combined to allow the search for signals with lower amplitude, and also to improve the frequency resolution. Bootstrap analysis was performed to distinguish between the most likely alias, and was also used to find errors on the peaks. In this method, the sampling pattern of the data is changed by creating a mock dataset where every point from the original dataset is randomised. This is done to efficiently destroy or weaken the aliases pattern.


\begin{table}
\begin{center}
 \begin{tabular}{llllll}
  \hline
  \hline
\textbf{Date} & \textbf{HJD} & \textbf{Telescope} & \textbf{Length} \\
& (-2455000) &  & (hours) \\
  \hline
100414 & 301 &  1.3m &  2.30 \\
100416 & 303   & 1.3m & 7.89 \\
100417 & 304   & 1.3m & 4.32 \\
100419 & 306   & 1.3m   & 4.72 \\
100424 & 311    & 1.3m & 5.42 \\
100425 & 312    & 1.3m  & 7.82 \\
100503 & 320  &  1.3m  & 7.92 \\
100504 & 321   & 1.3m  & 8.26 \\
100506 & 323   & 1.3m  & 8.40 \\
100507 & 324   & 1.3m  & 3.50 \\
100508 & 325   & 1.3m  & 5.47 \\
100528 & 345   & 2.4m  & 4.97 \\
100529 & 346   & 2.4m  & 8.11 \\
\hline
  \hline
\end{tabular}
 \caption{Summary observing log for SDSS J1457+51. Data was \newline obtained at the MDM observatory during April and May 2010.}
 \label{tab:obs1} 
\end{center}
\end{table}


    
\begin{figure*}
\centering
\subfigure[The mean Lomb-Scargle periodogram for J1457, from the six nights of best quality. The orbital period at 18.48 c\,d$^{-1}$ ($\omega_{\text{o}}$, 77.9 min) and its first harmonic ($2\omega_{\text{o}}$) are plotted as solid lines.] 
{
    \label{fig:s_avpow}
    \includegraphics[width=7.75cm]{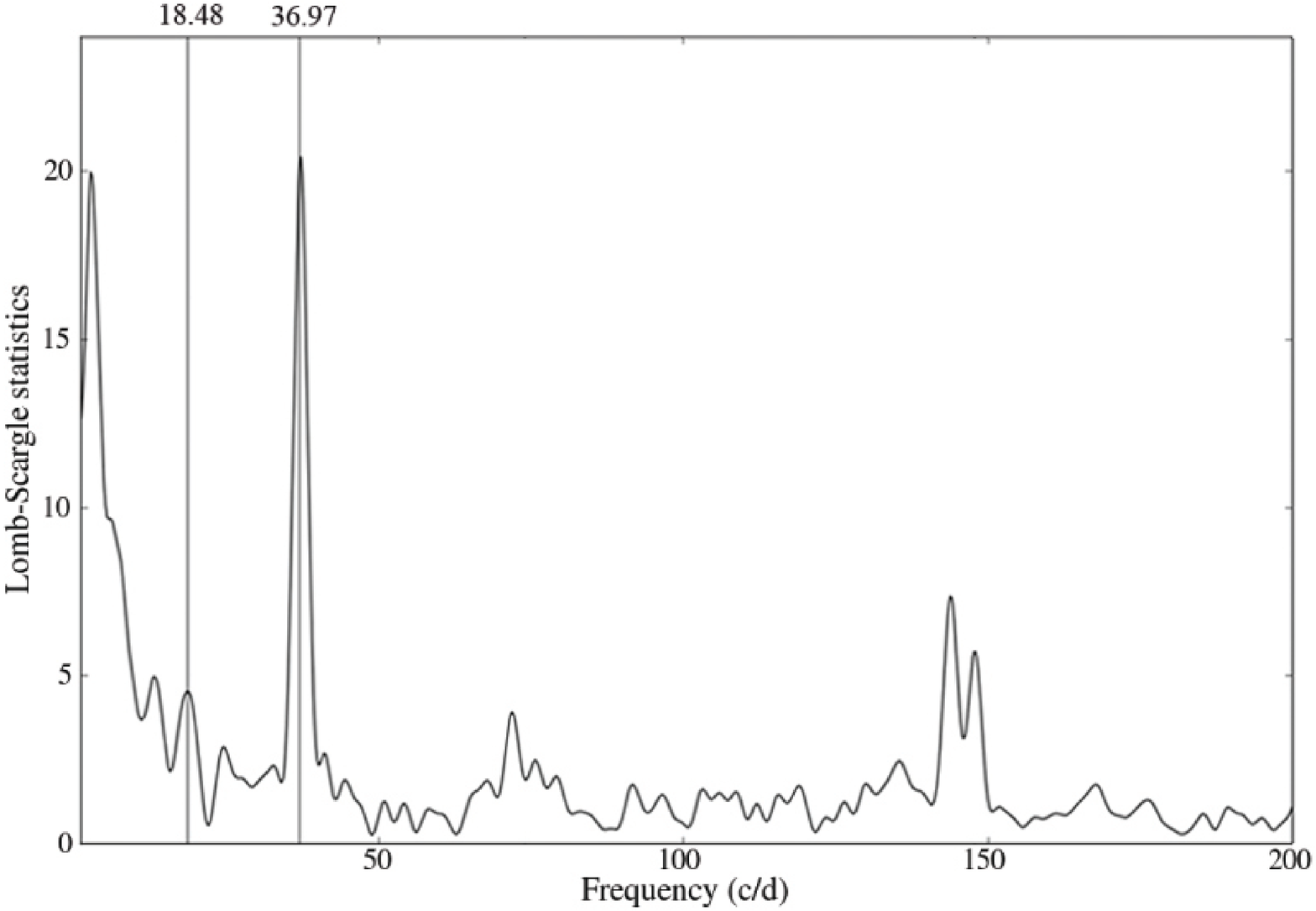}
}
\hspace{0.2cm}
\subfigure[Normalised and smoothed light curve of J1547 from one sample night. A model light-curve constructed from the four strongest periods (including $\omega_{\text{o}}$ and $2\omega_{\text{o}}$) found in Figure~\ref{fig:s_avpow}, is plotted together with the data.] 
{
    \label{fig:s_light}
    \includegraphics[width=7.12cm]{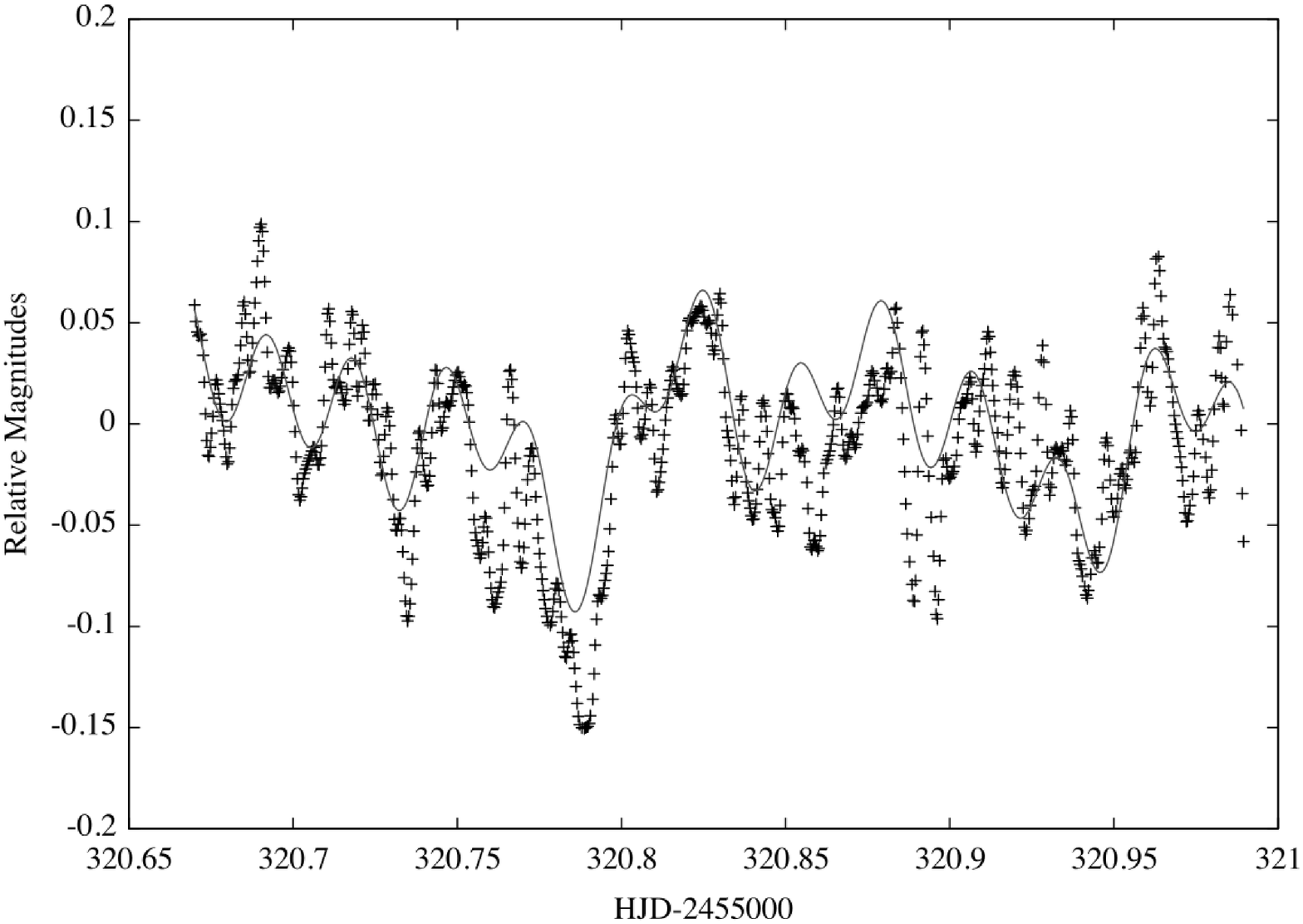}
}
\caption{}
\label{fig:sub} 
\end{figure*}


\subsection{Light Curve and average Power Spectrum}

The mean power spectrum, averaged over the six best nights is shown in Figure~\ref{fig:s_avpow}. The 18.48 c\,d$^{-1}$ (77.9 min) and 36.97 c\,d$^{-1}$ (38.9 min) signals are almost certainly the orbital frequency ($\omega_{\text{o}}$) and its first harmonic ($2\omega_{\text{o}}$). This kind of variation at $2\omega_{\text{o}}$ is commonly seen in the orbital light curves of CVs with low accretion rates (for instance in WZ Sge; see~\citealt{2002PASP..114..721P}). We find that the signals in the range 142 c\,d$^{-1}$ -- 148 c\,d$^{-1}$ ($\approx$ 10 min) are non-commensurate with the orbital frequency. These peaks vary slightly in period and amplitude, when present at all in the nightly power spectrum. The complex structure around them indicate either an unresolved fine structure, or periods varying from night to night, and is discussed in detail below. Also, during a few nights, peaks were found at 135 c\,d$^{-1}$ (10.7 min) and 72 c\,d$^{-1}$ (20 min), which are also non-commensurate with the orbital frequency. The lower region of the power spectrum show strong peaks at 4 c\,d$^{-1}$ -- 6 c\,d$^{-1}$ (4 -- 6 hours), corresponding to the typical length of a nightly observing run. The unit c\,d$^{-1}$ is used throughout the paper since it is the natural unit for the sampling pattern of the multi-day light curves. Also, it clearly shows the natural daily alias pattern.  

The average power spectrum in Figure~\ref{fig:s_avpow} has low resolution since each night is less than 8 hours long. The power spectrum of a spliced light curve spanning several nights (the coherent power spectrum) is in principle better, since the resolution is always near 0.1\,N$^{-1}$ c\,d$^{-1}$, where N is the duration in days. It does, however, make the assumption that a candidate periodic signal is constant in period, phase, and amplitude over the duration of the observation. Power spectra will be difficult or impossible to interpret correctly when this assumption is grossly violated.

Figure~\ref{fig:s_light} shows the normalised and smoothed light curve for a sample night. A model light curve constructed from the four strongest peaks found in the power spectrum that night (including both $\omega_{\text{o}}$ and $2\omega_{\text{o}}$), is plotted on top of the smoothed light curve. Peaks found at higher frequencies than 40 c\,d$^{-1}$ are not represented in the model light curve. During one of the observing nights, we obtained multicolour data and found the brightness of of the star to be V = 19.2 $\pm$ 0.2. The mean brightness on each night was constant within the measurement error of 0.03 mag. Flickering, an essentially universal feature of CV light-curves, was very low. This together with the fact that absorption lines are seen in the optical spectrum implies that the total light, 4000\,\AA\, -- \,7000\,\AA\, is dominated by the white dwarf. Measurements of the absorption line depth of the hydrogen lines indicate that probably no more than half of the total light comes from accretion processes.

\begin{figure*}
\centering
\subfigure[The power spectrum of J1457 from 11 nights, showing the orbital signal $\omega_{\text{o}}$ and its first harmonic, $2\omega_{\text{o}}$. A zoom of the region around $\omega_{\text{o}}$ is plotted on top.] 
{
    \label{fig:s_poworb}
    \includegraphics[width=7.15cm]{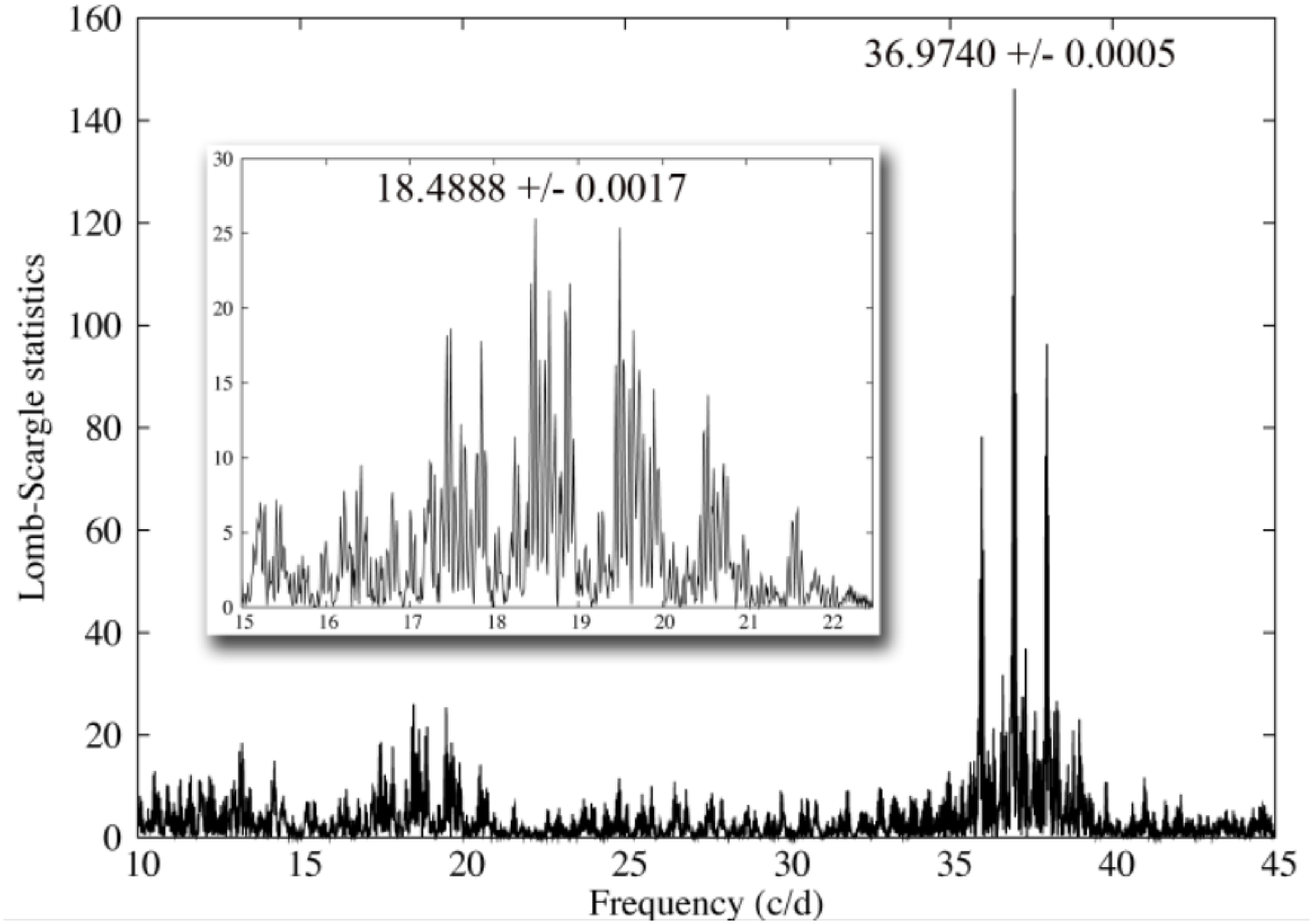}
}
\hspace{0.2cm}
\subfigure[A model power spectrum of J1457 constructed from two sine waves at $\omega_{\text{o}}$ and $2\omega_{\text{o}}$, using the same sampling pattern as in Figure~\ref{fig:s_poworb}.] 
{
    \label{fig:s_poworbmod}
    \includegraphics[width=7.15cm]{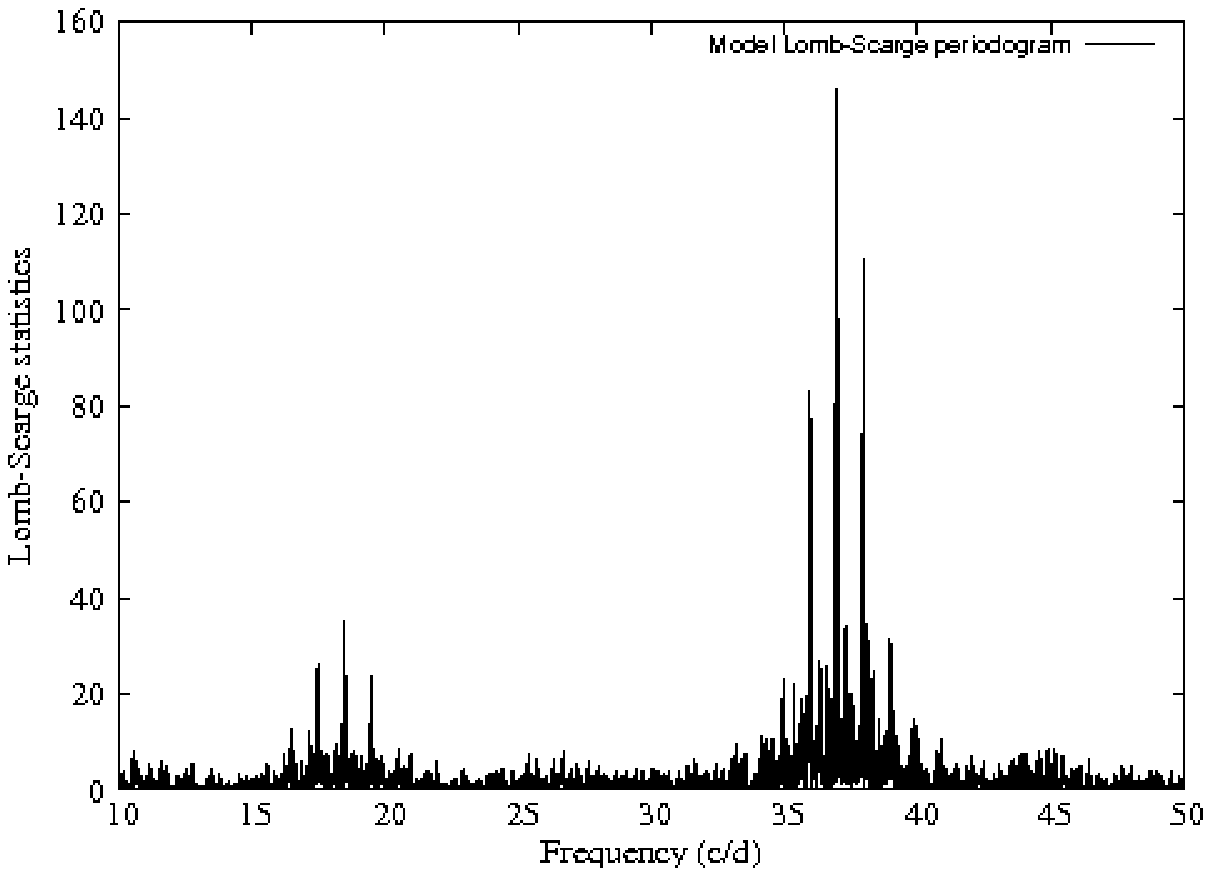}
}
\caption{}
\label{fig:sub} 
\end{figure*}


\begin{figure}
\includegraphics[width=7.5cm]{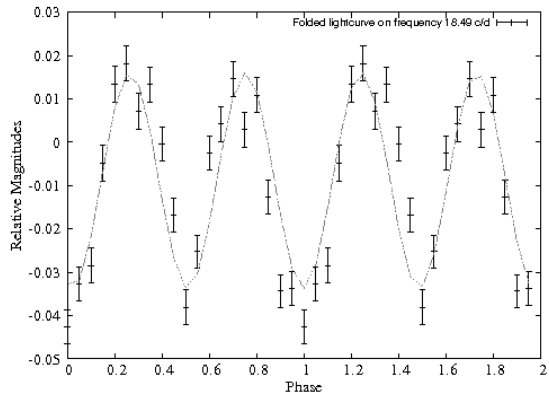}
\caption{Light curve of J1457 folded on the $\omega_{\text{o}}$ frequency, showing a double-humped orbital wave.} 
 \label{fig:fold18} 
\end{figure}


\subsection{The Orbital Signal}

A dominant, stable feature at $\approx$ 37 c\,d$^{-1}$ (38 min) is always present in the nightly power spectra. In four of the nights, a weaker but stable signal is also present at half that frequency,  $\approx$ 18.5 c\,d$^{-1}$  (78 min). We interpret these two signals as the orbital frequency and its first harmonic, $\omega_{\text{o}}$ and $2\omega_{\text{o}}$. As mentioned above, double-humped orbital waves are quite common among CVs, especially those of very low luminosity (for instance in WZ Sge and AL Com). A power spectrum composed of 11 nights, spanning 45 days, yields $\omega_{\text{o}} = 18.4888 \pm 0.0017$ c\,d$^{-1}$ and $2\omega_{\text{o}} = 36.9740 \pm 0.0005$ c\,d$^{-1}$. Errors are calculated from bootstrap simulations as described in Section~\ref{obs}.

Figure~\ref{fig:s_poworb} shows the low-frequency portion of the full 11 night power spectrum. A zoom of the region around the orbital frequency is plotted on top. A model power spectrum constructed from two artificial sinusoids at $\omega_{\text{o}}$ and $2\omega_{\text{o}}$, using the exact same sampling as for the original dataset, is shown in Figure~\ref{fig:s_poworbmod}. When comparing model versus data, we find the surrounding picket-fence pattern similar in structure and height. This implies that the orbital signal indeed maintains an essentially constant amplitude and phase. In Figure~\ref{fig:fold18}, data from all 11 nights are folded onto the orbital frequency, showing the double-humped orbital wave at $\omega_{\text{o}}$ and $2\omega_{\text{o}}$.  
 
The lower frequency range of the power spectrum was further investigated to rule out the possibility of signals hiding in the noise (see Section~\ref{superh} for the case of BW Sculptoris). The power spectrum was cleaned from the strongest signals at $\omega_{\text{o}}$, $2\omega_{\text{o}}$ and also from the high-amplitude peaks between 4 c\,d$^{-1}$ -- 6 c\,d$^{-1}$. However, no additional peak was found in this region, or in the vicinity of the orbital period.

 \begin{table*}
\begin{center}
 \begin{tabular}{lllllll}
  \hline
  \hline
\textbf{Star} & \textbf{Frequency} & \textbf{Period} & \textbf{Amplitude} & \textbf{Comments}  \\
 & (c\,d$^{-1}$) & (min) & (mmag) &   \\
 \hline
 J1457 & $18.4888 \pm 0.0017$  & 77.92 & 12.0 & Orbital period ($\omega_{\text{o}}$) \\      
& $36.9740 \pm 0.0005$ & 38.95 & 23.9 & $2\omega_{\text{o}}$ \\
& 71.9 (nightly mean error: 0.5) & 20.0 & 12.0 & NRP, low amplitude, non-stable \\ 
& 135.2/144.3/147.9 (nightly mean error: 0.8) & 10.7/9.9/9.7 & 3.9/12.7/7.6 & NRP, non-stable \\
\hline
BW Scl & $\sim$ 16.5 & 87.27 & $\sim$ 50 & Quiescent superhump, non-stable \\
 & 18.40811 $ \pm$ 0.03  & 78.23 & 16& $\omega_{\text{o}}$  \\ 
& 32.98 $\pm$ 0.01 & 43.66 & $\sim$ 15& Harmonic of quiescent superhump \\
& 36.81622 $ \pm$ 0.03  & 39.11 & 55& $2\omega_{\text{o}}$  \\
&  69.55 $\pm$ 0.03 & 20.70 &$\sim$ 25& NRP, $\omega_{1}$, non-stable  (amplitude from 2009)\\
& 103.55 $\pm$ 0.03  & 13.90 & $\sim$ 20& $\omega_{2}$ - $2\omega_{\text{o}}$, non-stable (amplitude from 2009) \\
& 140.37 $\pm$ 0.03 & 10.26 & $\sim$ 26& NRP, $\omega_{2}$, non-stable (amplitude from 2009) \\
& 121.99 $\pm$ 0.03  & 11.80 &$\sim$ 23.5 & $\omega_{2}$ - $\omega_{\text{o}}$, non-stable (amplitude from 2009) \\
& 153.0 $\pm$ 0.5 & 9.4 & 9.5& periodic signal (amplitude from 2001) \\
& 307.0 $\pm$ 0.5 & 4.7 & 8.5 & Harmonic of signal at 153 c\,d$^{-1}$, transient \\
  \hline
\hline
\end{tabular}
 \caption{Summary of frequencies found in SDSS J1457+51 and BW Sculptoris.}
\label{tab:freq} 
\end{center}
\end{table*}

\subsection{High-Frequency Power Excess}

The complex range of signals spanning 142 c\,d$^{-1}$ -- 148 c\,d$^{-1}$ ($\approx$ 10 min) moves slightly in frequency and is non-commensurate with the orbital frequency. In addition, during five of the observing nights a broad, low-amplitude peak appeared at 135 c\,d$^{-1}$ (10.7 min) along with a signal at 72 c\,d$^{-1}$ (20.0 min), neither of which are of orbital origin.  

With the aim to study these signals in more detail, a coherent power spectrum was constructed from adding consecutive nights together. However, this did not produce clean signals, even though the nightly power-spectrum windows were always clean. The broadness and complexity of the signals indicate a slight shift in amplitude and frequency on the time scale of a few nights, and/or an internal fine structure unresolved by our observations. Therefore, analysis was performed on power spectra from separate nights in comparison with the overall mean power spectrum.  

\begin{figure}
\includegraphics[width=7.5cm]{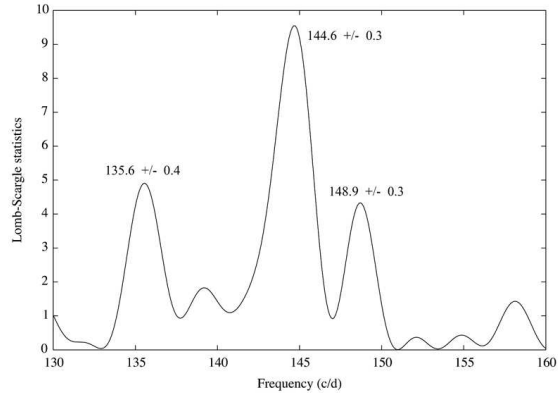}
\caption{Power spectrum of J1457 from one single night (HJD=2455320). Peaks were seen simultaneously at 135.6 c\,d$^{-1}$, 144.6 c\,d$^{-1}$ and 148.9 c\,d$^{-1}$.} 
 \label{fig:pow320} 
\end{figure}

When combining our sparse data collected over 43 nights, there is a broad power excess around the frequencies, 135 c\,d$^{-1}$, 144 c\,d$^{-1}$ and 148 c\,d$^{-1}$. This splitting is evident also in the mean spectrum shown in Figure~\ref{fig:s_avpow}, and is always seen when combining nights from the start and the end of the campaign. When studying the nightly power-spectrum in this range, there was (in general) only one peak present at the time. However, during one observing night, three peaks were seen simultaneously at approximately these frequencies (Figure~\ref{fig:pow320}). The nightly mean error for any peak appearing at 135 c\,d$^{-1}$ -- 148 c\,d$^{-1}$, is about 0.8 c\,d$^{-1}$.


\begin{figure}
\includegraphics[width=8.5cm]{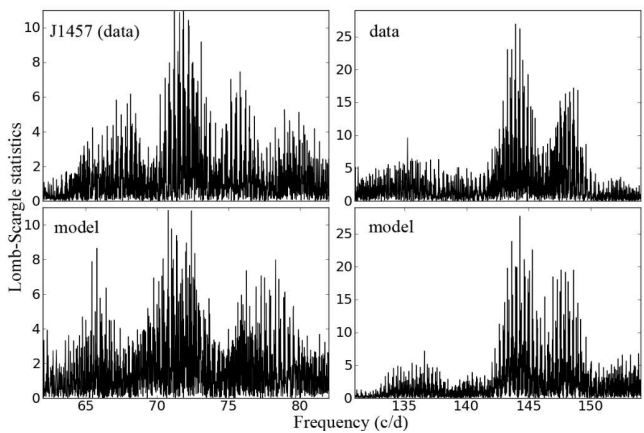}
\caption{The two top panels show the power spectrum of J1457 from the same set of nights shown in the mean spectrum (see Figure~\ref{fig:s_avpow}). The two bottom frames show models that was constructed from the same sampling pattern as the data. The following frequencies was included in the model; 18.48 c\,d$^{-1}$ ($\omega_{\text{o}}$), 36.97 c\,d$^{-1}$ ($2\omega_{\text{o}}$), 71.9 c\,d$^{-1}$, 135.2 c\,d$^{-1}$, 144.3 c\,d$^{-1}$ and 147.9 c\,d$^{-1}$. All main signals along with their line widths seen in the data can be reproduced fairly well by the model, indicating that there is power excess at these frequencies.} 
 \label{fig:s_pow_best2} 
\end{figure}

The upper two frames in Figure~\ref{fig:s_pow_best2} show the combined power spectra of the six nights of best quality. The bottom two panels show model power-spectra constructed from sine waves at the peak frequencies found in the combined six-night dataset, at 18.48 c\,d$^{-1}$ ($\omega_{\text{o}}$), 36.97 c\,d$^{-1}$ ($2\omega_{\text{o}}$), 71.9 c\,d$^{-1}$, 135.2 c\,d$^{-1}$, 144.3 c\,d$^{-1}$ and 147.9 c\,d$^{-1}$, using the same sampling pattern as the data. The model is able to re-construct the overall appearance and widths of the signals seen in the data reasonably well, indicating that there is power excess at these frequencies. If excluding any of the frequencies from the model, the data power spectrum cannot be reproduced. We note that 71.9 c\,d$^{-1}$ is about half that of 144.3 c\,d$^{-1}$, but not exactly in a 2:1 ratio, indicating that the signals are not constant in amplitude and phase (see Section~\ref{2009} for the case of BW Sculptoris). For a complete summary of the frequency analysis, see Table~\ref{tab:freq}. 


    
\section{BW Sculptoris}
 
BW Sculptoris (hereafter BW Scl) is a 16th magnitude blue star which was found to
coincide with RXJ2353-0-3852 in the Rosat bright-source catalogue, and
then identified as a cataclysmic variable by~\cite{1997A&A...318..134A}.~\cite{1997A&A...324L..57A} independently discovered the star as a blue object in the Hamburg/ESO
survey for bright QSOs. These two studies established the very short
orbital period of 78 minutes. In addition to broad and doubled H and He
emission lines, BW Scl also shows very broad Balmer and Lyman absorptions,
signifying the presence of a white dwarf of modest temperature ($\sim 15000$ K;
~\citealt{2005ApJ...629..451G}). If roughly half of the visual light comes
from such a white dwarf, then the white dwarf has $V=17.3$ and $M_{V}\sim 12$,
implying a distance of only $\sim 110$ pc. This also agrees with the large
proper motion found in the USNO catalogue (105 ms$\,$yr$^{-1}$, \citealt{2004AJ....127.3060G}). These considerations (a nearby star of very short $P_{\textbf{orb}}$), and the possibility to study the underlying white dwarf, motivated us to carry out campaigns of time-series photometry nearly every year since 1999.

\subsection{Observations}
 
In total, BW Scl was observed for about 1000 hours spread over about 200 nights, mainly using the globally distributed telescopes of the Center for Backyard Astrophysics (CBA: \citealt{1993ApJ...417..298S, patterson_1998}). A summary observing log is presented in Table~\ref{tab:obs2}. To maximise the signal and optimise the search for periodic features, usually no filter or a very broad filter, 4000\,\AA\, --\, 7000\,\AA\,was used. Occasional runs were obtained in V and I bandpasses to provide a rough calibration, and to verify that the periodic features in the light curve are indeed broadband signals. The smaller (25 cm -- 35 cm) telescopes generally used the star GSC 8015-671 as a comparison, while the larger (91 cm) telescopes used USNO 0450-40780391, a nearby 16th magnitude star. These comparison stars can be considered to be constant. The clear and broadband filters permit only a rough calibration, but BW Scl remained within $\sim 0.3$ mag of $V=16.6 $ throughout the campaign. In terms of instrumental magnitude, limits on night-to-night variability within each season are more stringent: typically $< 0.05$ mag, and always $< 0.1$ mag. This degree of constancy is remarkable for a cataclysmic variable, and is probably due to the WD's large contribution to the light in the optical. 
    
In order to study the periodic behaviour, we always tried to obtain photometry densely distributed in time, preferably with contribution from telescopes widely spaced in longitude (in order to solve problems associated with daily aliases). Most of the analysis below is based on long time series from stations in New Zealand, South Africa, and Chile, and hence not afflicted by aliasing problems.

 \begin{table}
  \begin{center}
 \begin{tabular}{llllll}
  \hline
  \hline
\textbf{Year} & \textbf{Spanned}  & \textbf{Observer} & \textbf{Telescope} & \textbf{Nights} \\
& (days) & &  & (hours) \\
 \hline
1999 &  13  &  Kemp & CTIO 91cm & 12/61\\      
2000 & 38 & McCormick & Farm Cove 25cm & 7/35\\
         &      & Rea    &     Nelson 35cm   &  2/9\\
2001  & 77  &    Rea   &           "     &         16/87\\
        &     & Kemp     &   CTIO 91cm       &   14/84\\
           &  & Woudt     &  SAAO 76cm           & 1/4\\
2002   & 0   &   Kemp      &       "               & 1/4\\
2004   & 7    &  Monard   &   Pretoria 25cm  & 7/38\\
2005  & 55    &  Rea       &  Nelson 35cm    & 16/66\\
          &  & Christie   & Auckland 35cm      & 12/48\\
        &  &  Retter/Liu   &                    & 6/32\\
    &   &  Monard     & Pretoria 25cm &  4/22\\
   &   &  Moorhouse   &                    & 3/10\\
2006  & 90   &   Rea        & Nelson 35cm  &  26/112\\
          &  & Monard     & Pretoria 35cm  & 3/21\\
&         &   McCormick  & Farm Cove 25cm & 5/15\\
2007  & 21  &    Rea        & Nelson 35cm   &  7/28\\
          &   & Richards  &  Melbourne 35cm     & 1/3\\
&         & Allen     &  Blenhem 41cm   & 1/3\\
2008  & 71   &   Monard   &   Pretoria 35cm & 22/88\\
&          &  Rea        & Nelson 35cm   & 13/90\\
2009  & 50  &    Rea         &     "             & 15/94\\
          &   & Monard      & Pretoria 35cm & 11/62\\
\hline
  \hline
\end{tabular}
 \caption{Summary observing log for BW Sculptoris.}
 \label{tab:obs2} 
\end{center}
\end{table}

\subsection{The 1999 Campaign}

\begin{figure}
\includegraphics[width=8.5cm]{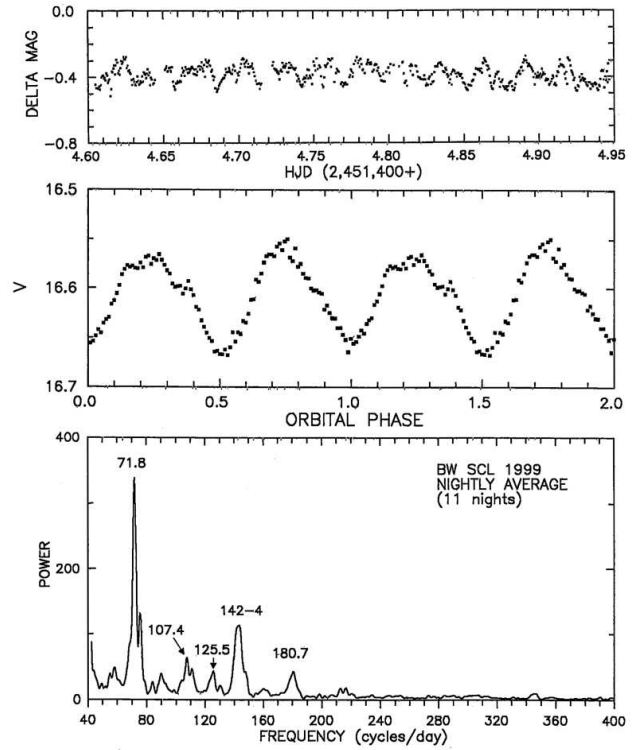}
\caption{BW Scl in 1999. The upper frame shows an 8-hour light curve, dominated by the orbital wave. The middle frame presents the season's mean orbital light curve, showing the double-humped waveform (the feature at $2\omega_{\text{o}}$ dominates all power spectra). The lowest frame shows the mean nightly power spectrum, averaged over the 11 best nights. Significant features are labelled with their frequency in c\,d$^{-1}$ (with an average error of $\pm$ 0.7).} 
 \label{fig:1} 
\end{figure}

The upper frame of Figure~\ref{fig:1} shows the light curve from one night during 1999. The general appearance is typical of all nights, as well as seen in previously published light curves (\citealt{1997A&A...324L..57A, 1997A&A...318..134A}). As those papers demonstrated, a 39-minute wave is always present, with a small even-odd asymmetry suggesting that the fundamental period is actually the double, 78 minutes (confirmed by spectroscopy). The middle frame shows the 1999 double-humped orbital light curve, similar to that of J1457 (see Figure~\ref{fig:fold18} above). Study of the seasonal timings yielded a 1995 -- 2009 ephemeris:
\\\\
Orbital maximum = HJD 2,450,032.182(3) + 0.05432392(2) E. 
\\\\
The average nightly power spectrum, the incoherent sum of the 11 nights of best quality, is shown in the bottom frame of Figure~\ref{fig:1}. Power excesses near 72 c\,d$^{-1}$ (20 min) and 143 c\,d$^{-1}$  (10 min) are evident. In order to study these higher-frequency signals in more detail, adjacent nights were added together, and the coherent power spectrum was constructed. However, even though the 72 c\,d$^{-1}$ and the 143 c\,d$^{-1}$ signals were always present, they were always broad, complex, and slightly variable in frequency (similar to J1457). This is usually a sign that the actual signals violate the assumptions of Fourier analysis: constancy in period, amplitude and phase. 
 

\begin{figure}
\includegraphics[width=8.5cm]{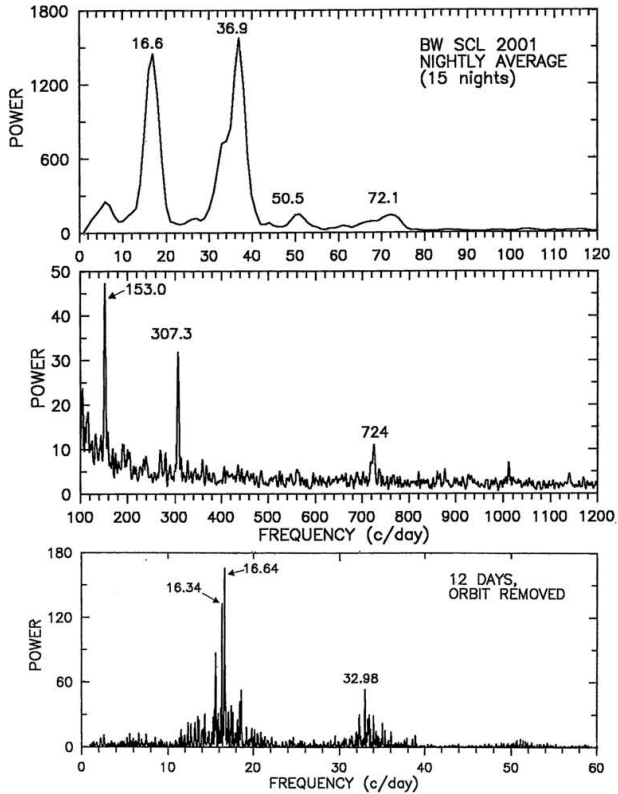}
\caption{BW Scl in 2001. The upper two frames show the mean nightly power spectrum, averaged over the 15 best nights. Features are labelled with their frequency in c\,d$^{-1}$ (with an average error of $\pm$ 0.7). The signal at 724 c\,d$^{-1}$, is likely to be caused by instrumental effects. The lowest frame shows the power spectrum of a particularly dense 12-night segment, after subtraction of the orbital waveform. A powerful signal near 16.5 c\,d$^{-1}$ is obvious, with a strong first harmonic. The formal error is $\sim$ 0.01 c\,d$^{-1}$, and each peak is about that wide. However, the power excess shows more structure and width than seen in the power-spectrum window (see Figure~\ref{fig:4} below), indicating some amplitude and/or phase modulation.} 
 \label{fig:2} 
\end{figure}


\subsection{The 2001 Campaign}
 
During 2001, observatories at three longitudes in Chile, New Zealand 
and South Africa, contributed with good nightly coverage of BW Scl. A very low flickering background is seen in the light curves of BW Scl, enabling detection of periodic signals as weak as $\sim 0.002$ magnitudes.

The upper frames of Figure~\ref{fig:2} shows the nightly mean power spectrum, with significant signals marked to an accuracy of 0.5 c\,d$^{-1}$. Both the orbital period and the signal at 72 c\,d$^{-1}$  (20 min) were present, along with two other signals at low frequency: a powerful signal at 16.6 c\,d$^{-1}$  (86.7 min), and a weak signal at 50.5 c\,d$^{-1}$ (28.5 min). At higher frequency, signals are detected at 153 c\,d$^{-1}$ (9.4 min), 307 c\,d$^{-1}$ (4.7 min), and 724 c\,d$^{-1}$ (2 min), though the latter is likely to be caused by instrumental effects. Many telescopes have worm gears which turn with a period of exactly 120 sidereal seconds, and this period was reported in research on many types of stars during 1960 -- 1990, i.e. during the photolectric-photometer era. CCDs are much less prone to this error. However, since 724 $\pm$ 2 c\,d$^{-1}$ corresponds to 120 sidereal seconds (to within the measurement error), we interpret the signal to be caused by this instrumental effect.   
 
We interpret the signal at 153 c\,d$^{-1}$ as a pulsation frequency, with a significant first harmonic. Examination of individual nights showed this signal to be somewhat transient, at least in amplitude. This behaviour was seen on 7 of the 15 good-quality nights. Since the orbital frequency is known precisely, and since its photometric signature is powerful and constant, we subtracted its first harmonic (and also the second harmonic when detected) from the central 12-night time-series, prior to analysis. The resultant power spectrum at low frequency is seen in the bottom panel of Figure~\ref{fig:2}. A weak signal appears at the orbital frequency, and stronger signals at 16.34/16.64 c\,d$^{-1}$ and 32.98 c\,d$^{-1}$ (with an accuracy of $\pm\,\,0.01$ c\,d$^{-1}$). It seems likely that these are superhump signals in the quiescent light curve. Specifically, we interpret the 16.34/16.64 pair as signifying an underlying frequency of $\sim 16.5$ c\,d$^{-1}$, as such splitting can be produced by amplitude and/or phase changes. The precession frequency $\Omega$ can be expressed as $16.50 = \omega_{o} - \Omega$, where $\omega_{o}$ is the orbital frequency (implying that $32.98 = 2(\omega_{o} - \Omega)$). The precession frequency $\Omega$ is then equal to 1.9 c\,d$^{-1}$.

The primary signal near 16.5 c\,d$^{-1}$ has been seen before. It was the dominant signal reported by~\cite{1997A&A...318..134A}, but was then discounted by~\cite{1997A&A...324L..57A} as probably the result of cycle-count error. The data presented here certainly has no ambiguity in cycle count, and reveals this signal very clearly. 



\begin{figure}
\includegraphics[width=8.5cm]{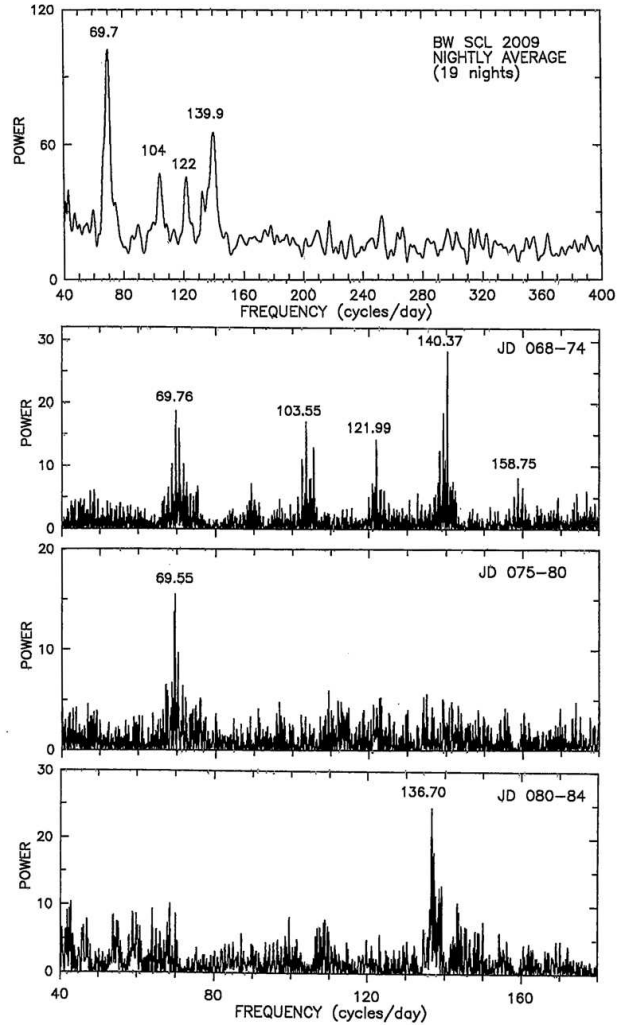}
\caption{ BW Scl in 2009. The top frame shows the mean nightly power spectrum, averaged over the 19 best nights. Features are labelled with their frequency in c\,d$^{-1}$ ($\pm$ 0.6). Other frames show power spectra of $\sim$ 5 day intervals with particularly dense coverage, with significant features labelled ($\pm$ 0.03). The most interesting features are the trio of weak satellites of $\omega_{2}$=140.37 c\,d$^{-1}$ (103.55 c\,d$^{-1}$ and 121.99 c\,d$^{-1}$ are displaced by exactly $\omega_{2}$ - $2\omega_{\text{o}}$ and $\omega_{2}$ - $\omega_{\text{o}}$).} 
 \label{fig:3} 
\end{figure}

 
\subsection{The 2009 Campaign}  \label{2009}

The year 2009 saw another intensive observing campaign on BW Scl. The upper frame of Figure~\ref{fig:3} shows the nightly power spectrum, averaged over the 19 nights of coverage. Signals near 70 c\,d$^{-1}$  (20.6 min) and 140 c\,d$^{-1}$  (10.3 min) are evident, with weaker signals near 104 c\,d$^{-1}$  (13.9 min) and 122 c\,d$^{-1}$  (11.8 min). In the three lower frames, we show intervals of particularly dense coverage, each with a nominal frequency resolution of $\pm\,\,0.04$ c\,d$^{-1}$. In these frames, the 70 c\,d$^{-1}$ and 140 c\,d$^{-1}$ signals show their variability in frequency and amplitude (similar to J1457). Such variability is characteristic of all our data. The second frame (JD 2,455,068 - 074) illustrates the following:

\begin{enumerate}
\item [1.] Although the 70 c\,d$^{-1}$ and 140 c\,d$^{-1}$ signal are related (when one moves
to slightly lower frequency, so does the other), the frequencies do not appear to be exactly in the ratio 2:1. Here, these frequencies are described as $\omega_{1}$ and $\omega_{2}$, with $\omega_{2} \sim 2\omega_{1}$.

\item [2.] The weaker signals flanking the 140 c\,d$^{-1}$ feature are displaced by exact
integer multiples of the orbital frequency. Thus the signals seen in JD 068-74 are $\omega_{1}$, $\omega_{2}$, ($\omega_{2} - \omega_{\text{o}}$) and ($\omega_{2} - 2\omega_{\text{o}}$). These orbital sidebands of $\omega_{2}$ are also visible in Figure~\ref{fig:1}, however such orbital sidebands was never seen of $\omega_{1}$. This reproduces the properties of the pulsations seen in SDSS J1507+52 (\citealt{2008PASP..120..510P}).
\end{enumerate}

\subsection{Power-Spectrum Window}
  
\begin{figure*}
\includegraphics[width=13cm]{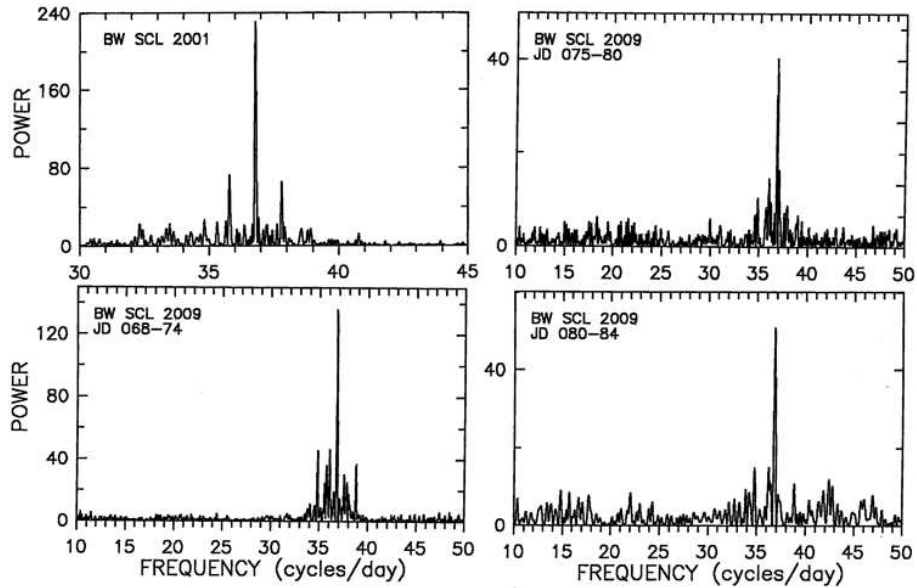}
\caption{BW Scl. Power spectra in the vicinity of $2\omega_{\text{o}}$ for the lowest frame in Figure~\ref{fig:2}, and the three lower frames in Figure~\ref{fig:3}. Since the $2\omega_{\text{o}}$ signal is essentially constant in frequency and amplitude, this is effectively the power-spectrum window for each observation. This implies that the nightly length of run and span of longitude is sufficient to exclude confusion by aliases.} 
 \label{fig:4} 
\end{figure*}

Finally, in Figure~\ref{fig:4} we show the power spectrum in the vicinity of $2\omega_{\text{o}}$, for each season. The $2\omega_{\text{o}}$ signal maintains constancy in phase and amplitude, and therefore acts effectively like a test signal whose power spectrum should be reproduced in structure by any coherent signal (constant in phase and amplitude). The central peak at exactly 36.816 c\,d$^{-1}$ (39.113 min) always dominates over its neighbours in the picket-fence pattern because of very long runs, and/or observations at several longitudes. This demonstrates that these campaigns are free from aliasing. Also, it is worth noting that peaks at frequencies greater than $2\omega_{\text{o}}$, are always slightly wider (for the single-night time series) than expected for a truly periodic signal, and always more complex (for the multiple-night time series). This is true for all our data. The underlying $\omega_{1}$ and $\omega_{2}$ signals either have an intrinsic complex structure, or vary strongly in amplitude and/or phase on short timescales (or both). In addition, the entire complex of signals near $\omega_{1}$ and $\omega_{2}$ moves in frequency by a few percent on timescales as short as $\sim10$ days.
  
\subsection{Superhumps in BW Scl?} \label{superh}

The obvious signal at 87 minutes in BW Scl, displaced by $\sim 11$ \% from $P_{\text{orb}}$ seen in the 2001 power spectrum (Figure~\ref{fig:2}), is a transient but repeating feature in the light curve. In our ten years of coverage, six campaigns were sufficiently extensive to reveal such a signal. This feature was always strong when it was detected, and moved slightly in frequency, even on timescales of a few days. This qualitatively describes a common superhump, which is a well-known feature in CV light-curves. However, there are three substantial differences:
\begin{enumerate}
\item [1.] Common superhumps are found in outburst states, typically near $M_{V} \sim 5$, not at quiescence when $M_{V} \sim 12$.
\item [2.] Common superhumps occur with fractional period excesses, $(P_{\text{sh}} - P_{\text{orb}})/P_{\text{orb}}$, near 3\%, not 11\%.
\item [3.] Common superhumps are more stable, wandering from a constant-period ephemeris on a timescale of 100 -- 200 cycles. The power-spectrum signal in BW Scl appears complex and broad, suggesting much lower coherence.
\end{enumerate}

A system with quiescent superhumps is a rare beast in the CV kingdom, but it is not unprecedented. In AL Com, a nearly identical signal was reported and extensively discussed by \cite{1996PASP..108..748P}. Something similar has also been reported in V455 And (\citealt{2005A&A...430..629A}), SDSS J0745+45 (\citealt{2010ApJ...710...64S}) and possibly SDSS J1238-03 (\citealt{2010ApJ...711..389A}). All these stars are period-bouncer candidates (systems that have already passed the minimum period; see Tables 3 and 5 of \citealt{2011MNRAS.tmp...27P}). These candidates were chosen due to their low donor-star mass (or low $q = M_{2}/M_{1}$). A possible account of how stars of very low $q$ might be able to manufacture quiescent superhumps has been given in~\cite{1996PASP..108..748P}, and includes the idea that a low donor mass implies a larger Roche lobe surrounding the disc. Weak tidal torques could then allow the quiescent disc to extend to the 2:1 orbital resonance, where an eccentric instability could drive a fast prograde precession (viz., at $\Omega$ = 1.9 c\,d$^{-1}$), resulting in a superhump with $\omega = \omega_{o} - \Omega$.


\section{Interpretations as Non-Radial Pulsations}

Both BW Scl and J1457 are quiescent CVs of very low accretion luminosity. Their spectrum show evidence of the primary WD, as do their light curves which contain rapid, non-commensurate signals. This is the general signature of the GW Lib stars, where the periodic signals are believed to represent the non-radial pulsations of the underlying WD. However, no proof of this has ever been found, not for GW Lib or any other of the 10 -- 15 members of the class. In order to explain these signals, two other hypotheses deserve consideration: first, quasi-periodic oscillations (QPOs) arising from the accretion disc, and second, the spin frequency of a magnetic WD (intermediate-polar model).

QPOs were first discovered, and named, for a broad excess of power seen during a dwarf-nova eruption of RU Pegasi (\citealt{1977ApJ...214..144P}). Since then, they have been commonly found with periods of 10 -- 20 minutes in the high-state light-curves of many nova-like variables (reviewed by \citealt{2004PASP..116..115W}). QPOs are typically seen as very broad peaks in the power spectrum ($\delta v/v \sim 0.5$; see for instance Figure 11 and 12 in~\citealt{2002PASP..114.1364P}), and typically have modulations of 1\% -- 3\% of an accretion disc in full outburst (for a light source with $M_{V}\sim 4$). The most famous example is the 20-minute signal in TT Arietis (which belongs to the class of VY Sculptoris stars). However, VY Sculptoris stars appear to be the among the most luminous CVs. This can be compared to the faint BW Scl, where the 70/140 c\,d$^{-1}$ signals have $\delta v/v \sim 0.01$ and are $\sim 1$\% modulations of a light source with $M_{V}\sim 12$. The peaks in J1457 have slightly higher $\delta v/v$, but on both counts, the QPO hypothesis earns no applause. It would have to be an essentially new kind of accretion-disc QPO.

A white dwarf spinning with a period of 20 minutes could explain the features of the signal seen at $\sim 70/140$ c\,d$^{-1}$ in both objects, the common appearance of the first harmonic, the occasional switch to a pure first harmonic (two-pole accretion), and the orbital sidebands seen in BW Scl (from amplitude modulations, and/or reprocessing from structures fixed in the orbital frame). But this hypothesis fails to account for the shifts in frequency exhibited by the $\omega_{1}$ and $\omega_{2}$ signals (the $\sim 5$\% wandering, e.g. in the range 68 c\,d$^{-1}$ -- 73 c\,d$^{-1}$ and 135 c\,d$^{-1}$ -- 153 c\,d$^{-1}$), the fact that $\omega_{2}$ is not exactly $2\omega_{1}$, and the intrinsic breadth (or fine structure) of both signals. These are profound inconsistencies.

Such considerations drive us back to the GW Lib model. Actually, we have studied all of the 10 -- 15 known class members with time-series photometry, and the resemblances to BW Scl and J1457 are substantial:
\begin{enumerate}
\item [1.] Low-amplitude and non-commensurate periodic signals in low-$\dot{M}$ CVs. 
\item [2.] Signals roughly constant over a few days, but somewhat transient in
    frequency and amplitude on longer timescales.
\item [3.] Signals frequently with strong first harmonics or quasi harmonics.
\item [4.] Signals sometimes with known or suspected fine structure.
\end{enumerate}
These resemblances, and the difficulties with alternative models, seem
sufficient to accept both BW Scl and J1457 as new members of the GW Lib class.     
     
\section{Summary}

\begin{enumerate}
\item [1.] Two more CVs of very low accretion-rates have shown rapid non-commensurate signals in quiescence, which makes them likely new members of the GW Lib class. Both J1457 and BW Scl show a complex spectrum with the main signals near 10 and 20 minutes.

\item [2.] The pulsation frequencies in both stars wander by a few percent on a timescale of days. In addition, the power-spectrum constructed from multiple-night light curves, show broad peaks, which might be due to the frequency wandering, or from intrinsic fine structure not resolved by our data (or due to a strong amplitude modulation).

\item [3.] BW Scl shows several peaks displaced from the main pulsation frequency $\omega_{2}$, by $\omega_{2} \pm \omega_{\text{o}}$ and $\omega_{2} \pm 2\omega_{\text{o}}$. Similar behaviour is evident in other GW Lib stars (SDSS J1507+52, V386 Ser and SDSS J1339+48). The origin of this phenomenon is still unknown. The rich pulsation spectrum makes BW Scl a good candidate for an intensive round-the-world time-series campaign with larger telescopes.

\item [4.] The orbital light curves of both stars show double-humped waves. From these waves, precise periods are found at $P_{\text{orb}} = 78.22639 \pm 0.00003$ minutes for BW Scl, and $77.885 \pm 0.007$ minutes for J1457. Similar systems displaying non-radial pulsations, such as SDSS J1339+4847, SDSS J0131-0901 and SDSS J0919+0857, all have orbital periods right at the same orbital period ($\approx$ 80 minutes). 

\item [5.] BW Scl sometimes shows a transient wave with a period, $P_{\text{sh}} = 87.27$ minutes, which is interpreted as a quiescent superhump. It thereby joins a small group of stars who manage a superhump at quiescence, all of which are likely to have very low mass ratios. This might arise from an eccentric instability at the 2:1 resonance in the disc.

\item [6.] The white-dwarf domination of the spectra in these stars suggests great faintness of the accretion light. This signifies a very low accretion rate, and both stars are likely to be very old CVs.
\end{enumerate}

\section*{Acknowledgments}   

Joe Patterson would like to acknowledge the support from NASA grant GO11621.03A, the Mt Cuba Astronomical Foundation and the NSF grant AST 0908363.


\bibliography{ref} 

\bsp

\label{lastpage}

\end{document}